# The effects of interfacial recombination and injection barrier on the electrical characteristics of perovskite solar cells

Lin Xing Shi [1, a], Zi Shuai Wang [2], Zengguang Huang [1], Wei E. I. Sha [3, b], Haoran Wang [1], Zhen Zhou [1]

[1] *School of Science, Huaihai Institute of Technology, Lianyungang, 222005, P. R. China*

[2] *Department of Electrical and Electronic Engineering, The University of Hong Kong, Pokfulam Road, Hong Kong*

[3] *College of Information Science & Electronic Engineering, Zhejiang University, Hangzhou 310027, P. R. China*

Charge carrier recombination in the perovskite solar cells (PSCs) has a deep influence on the electrical performance, such as open circuit voltage, short circuit current, fill factor and ultimately power conversion efficiency. The impacts of injection barrier, recombination channels, doping properties of carrier transport layers and light intensity on the performance of PSCs are theoretically investigated by drift-diffusion model in this work. The results indicate that due to the injection barrier at the interfaces of perovskite and carrier transport layer, the accumulated carriers modify the electric field distribution throughout the PSCs. Thus, a zero electric field is generated at a specific applied voltage, with greatly increases the interfacial recombination, resulting in a local kink of current density-voltage (*J-V*) curve. This work provides an effective strategy to improve the efficiency of PSCs by pertinently reducing both the injection barrier and interfacial recombination.

## I. Introduction

Despite high power conversion efficiency (PCE), over 20%, of organic-inorganic lead halide perovskite solar cells (PSCs) has been reported in recent years, arising from a high absorption coefficient, high carrier mobilities, and long charge carrier diffusion lengths[1–3], the current density-voltage (*J-V*) responses represent an anomalous hysteresis[4–6] and distortion. It suffers from a challenge to get the actual PCE of PSCs. Chen[7] reviewed the recent progress on the investigation of the origin of *J-V* hysteresis behavior in PSCs: slow transient capacitive current[8], dynamic trapping and detrapping processes[9], and band bending due to ion migrations[10] or ferroelectric polarization[11]. To describe the operations of PSCs, some numerical device modelings based on non-linear Poisson and drift-diffusion equations without[12–14] or with[15–17] ion migration have been developed. According to the transient simulation of the photovoltage and the photocurrent, Calado[15] concluded that hysteresis requires the combination of both the mobile ionic charge and the recombination near the perovskite-contact interfaces. Ion migrations modified the net built-in electric field throughout the PSCs and the trap-assisted recombination at the perovskite charge collection layer interface. Consequently, with ion migration and interfacial recombination, PSCs exhibit the S-shaped concave deformation of their J-V characteristics at forward sweep. The charge transport restrictions and the interfacial recombination are regarded to be mainly responsible for the S-shaped kink.

[a] slxopt@hotmail.com
[b] weisha@zju.edu.cn

It is well known that the recombination of charge carriers in the PSCs will reduce not only the fill factor (*FF*) but also the open-circuit voltage ($V_{OC}$). Most of the structures of PSCs are composed of electron transport layer (ETL), perovskite absorber layer and hole transport layer (HTL). Apart from grain boundaries which act as bulk recombination in perovskite absorber layer, various film interfaces recombination[18–20] from ETL/perovskite or perovskite/HTL could play different roles in the electrical performance of PSCs. Furthermore, in addition to the interfacial recombination, an injection barrier arises at the interface as well, as shown in Fig. 1(b), which results in charge carrier accumulation. The combination of large enough injection barrier and interfacial recombination could also produce a ‘local’ kink of the J-V curves. It is different from the S-shaped kink which is a 'global' kink. In view of these statements, in this work, we will comprehensively investigate the impacts of injection barrier, recombination channels, doping properties and light intensity on the electrical performance of PSCs. Particularly, the physical origins of the corresponding local kink characteristic of PSCs are discussed and understood by drift-diffusion model.

## II. Device structure and model

### A. Device structure

The solar cell device to be investigated has an n-i-p device structure, where n is the doped $TiO_2$ acting as ETL, i is the perovskite ($CH_3NH_3PbI_3$) absorber layer, and p is the doped spiro-OMeTAD acting as HTL[21]. The devices employ FTO and gold (Au) as the cathode and anode, respectively[17]. Figure 1(a) shows the device configuration under study and three different recombination mechanisms are to be studied, i.e., Case 1: the bulk recombination from the trap-assisted recombination at grain boundaries, denoted as 'Bulk'; Case 2: the ETL/perovskite interfacial recombination from the trap-assisted recombination at its interface, denoted as 'Top'; Case 3: the perovskite/HTL interfacial recombination from the trap-assisted recombination at its interface, denoted as 'Bottom'. The layer thicknesses of the PSCs are in consistent with the parameters defined in Ref. 17. The schematic illustration of the energy band diagram of PSCs is shown in Fig. 1(b).

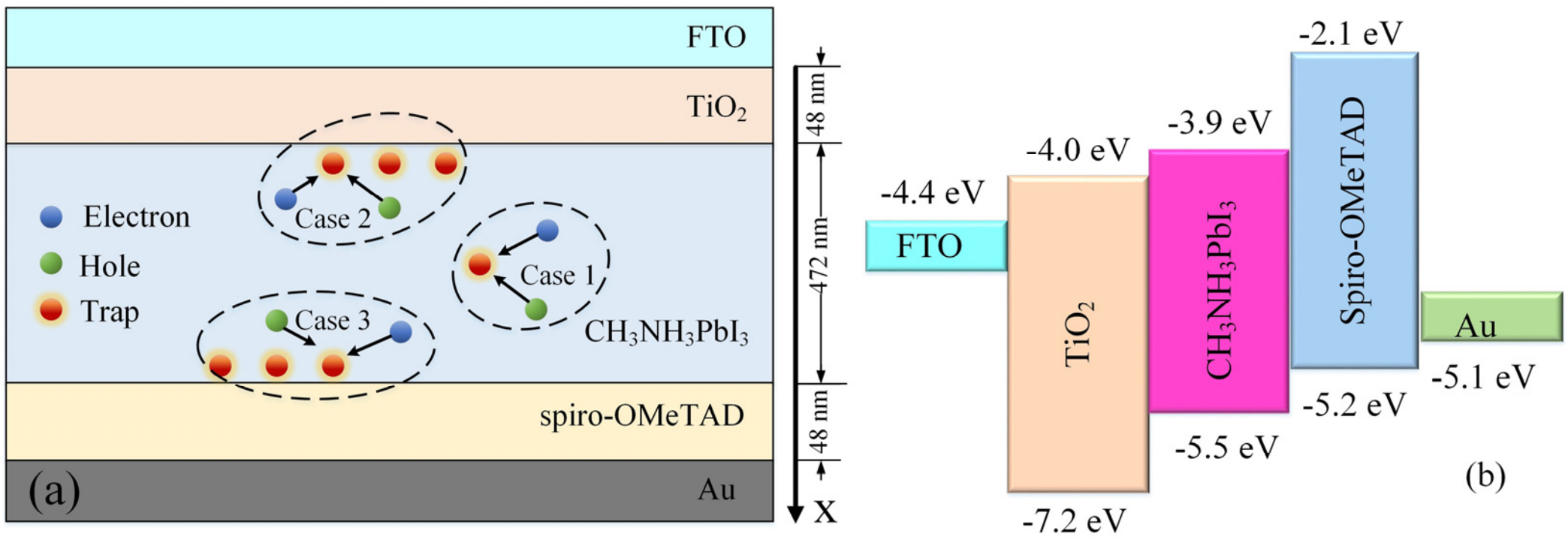

FIG. 1. (a) Device structure of perovskite solar cell under study. Case 1: bulk recombination at grain boundaries, denoted as 'Bulk'; Case 2: ETL/perovskite interfacial recombination at its interface, denoted as 'Top'; Case 3: perovskite/HTL interfacial recombination at its interface, denoted as 'Bottom'. (b) Schematic illustration of the energy band diagram of perovskite solar cell to be modeled.

**B. Simulation model**

The device model is based on the non-linear Poisson, drift-diffusion and continuity equations for electrons and holes throughout the device in one dimension[22]. The transport of charge carriers is governed by the electrically induced drift and diffusion for electrons and holes,

$$\frac{\partial^2 V}{\partial x^2} = -\frac{q}{\varepsilon}(p - n - N_A^- + N_D^+) \tag{1}$$

$$\frac{\partial n}{\partial t} = \frac{\partial}{\partial x}\left(\mu_n nE + D_n \frac{\partial n}{\partial x}\right) + G_n - R_n \tag{2}$$

$$\frac{\partial p}{\partial t} = -\frac{\partial}{\partial x}\left(\mu_p pE - D_p \frac{\partial p}{\partial x}\right) + G_p - R_p \tag{3}$$

$$J_n = q\mu_n nE + qD_n \frac{\partial n}{\partial x} \tag{4}$$

$$J_p = q\mu_p pE - qD_p \frac{\partial p}{\partial x} \tag{5}$$

Where $V$ is the electrostatic potential, $q$ is the positive electron charge, $\varepsilon$ is the dielectric permittivity, $n$ and $p$ are the densities of the electrons and holes, $N_A^-$ and $N_D^+$ are the ionized p-type and n-type doping. The doping levels are constant in the ETL and HTL, and zero in the perovskite layer. $E$ is the built-in electric field, $\mu_n$ and $\mu_p$ are the mobility of electrons and holes, respectively. $D_n = \mu_n(k_B T/q)$ and $D_p = \mu_p(k_B T/q)$ are the diffusion coefficients of electrons and holes, respectively, where $k_B$ and $T$ are the Boltzmann constant and Kelvin temperature. $G_n$ and $G_p$ are the generation rate of electrons and holes, respectively. The charge carrier generation profile under illumination throughout the device, calculated by using the Finite-Difference Time-Domain (FDTD) method[23], is depicted in Fig. S1. $R_n$ and $R_p$ are the recombination rate of electrons and holes, respectively. Here, it is provided that the trap-assisted recombination $(R_S)$[17] is dominant in the PSCs. $J_n$ and $J_p$ are electron and hole current densities, respectively.

The boundary condition of the electrostatic potential at the electrodes is

$$V = V_a - W_m/q \tag{6}$$

where $V_a$ is the externally applied voltage and $W_m$ is the work function of the electrode.

The charge carrier densities (boundary conditions) at the Schottky contracts are given by

$$\text{Cathode}\begin{cases} n(0) = N_c \exp\left(\frac{-q\emptyset_n}{k_B T}\right) \\ p(0) = N_v \exp\left(\frac{-E_g + q\emptyset_n}{k_B T}\right) \end{cases} \tag{7}$$

$$\text{Anode}\begin{cases} n(L) = N_c \exp\left(\frac{-E_g + q\emptyset_p}{k_B T}\right) \\ p(L) = N_v \exp\left(\frac{-q\emptyset_p}{k_B T}\right) \end{cases} \tag{8}$$

where $N_c$ and $N_v$ are the effective density of states for charge carrier transport materials. $\emptyset_n$ and $\emptyset_p$ are injection barriers for the cathode and anode, respectively.

The computational method[22,24,25] used in solving the non-linear Poisson and drift-diffusion equations is shown in the Supplementary Information (SI). The general device model parameters[13,17] for $TiO_2$, $CH_3NH_3PbI_3$ and spiro-OMeTAD are listed in Table I.

To investigate the impacts of trap-assisted interfacial recombination on the performance of PSCs, the traps are located in a 5 nm thick interface recombination region[13] between ETL and perovskite (Case 2) and, perovskite and HTL (Case 3), respectively. Suppose that the traps density for interfacial recombination is $N_t$, for equal comparison, then the traps density for bulk recombination is $N_t \times L_i/L_p$. Here, $L_p$ is the thickness of perovskite and $L_i$ is the thickness of interface recombination region.

**Table I**
General Device Model Parameters

| Property | Unit | $TiO_2$ | Perovskite | spiro-OMeTAD |
|---|---|---|---|---|
| $\boldsymbol{L}$ | nm | 48 | 472 | 48 |
| $\boldsymbol{E_g}$ | eV | 3.2 | 1.6 | 3.1 |
| $\boldsymbol{\chi}$ | eV | 4.0 | 3.9 | 2.1 |
| $\boldsymbol{N_{c,v}}$ | $cm^{-3}$ | $10^{20}$ | $10^{20}$ | $10^{20}$ |
| $\boldsymbol{\varepsilon_r}$ | | 80 | 6.5 | 3.0 |
| $\boldsymbol{N_D^+}$ | $cm^{-3}$ | $8\times10^{17}$ | 0 | 0 |
| $\boldsymbol{N_A^-}$ | $cm^{-3}$ | 0 | 0 | $8\times10^{17}$ |
| $\boldsymbol{\mu_n}$ | $cm^{-3}V^{-1}s^{-1}$ | 2 | 2 | 0.02 |
| $\boldsymbol{\mu_p}$ | $cm^{-3}V^{-1}s^{-1}$ | 0.02 | 2 | 2 |
| $\boldsymbol{N_t}$ | $cm^{-3}$ | 0 | $10^{14}\sim10^{17}$ | 0 |
| $\boldsymbol{C_{n,p}}$ | $cm^{-3}s^{-1}$ | 0 | $10^{-7}$ | 0 |

Sun et al.[26] prepared less-crystallized nanoporous $PbI_2$ (ln-$PbI_2$) based perovskite solar cells with the solid-state reaction method at a low temperature. Compared with compact $PbI_2$ (c-$PbI_2$) counterparts, it delivered much higher PCE resulting from decreased non-radiative defects, as shown in Fig. 4(b) in their paper. However, the *J-V* curve of ln-$PbI_2$ based PSCs presents a little distortion when the applied voltage is near 0.8 V. Thus, we perform a simulation and fit the simulated data to the experimental data. The *J-V* curves of numerical and experimental results are shown in Fig. 2, which verify our model. We find that the combination of the injection barrier (between perovskite and carrier transport layers) and the trap-assisted bulk and interfacial recombination could near-perfectly fit the experimental data. The injection barrier between perovskite and ETL is 0.1eV and the one between perovskite and HTL is 0.47eV. The trap density is $5\times10^{11}cm^{-3}$, $7.6\times10^{16}cm^{-3}$ and $7.6\times10^{16}cm^{-3}$ for bulk, top and bottom interfacial recombination, respectively.

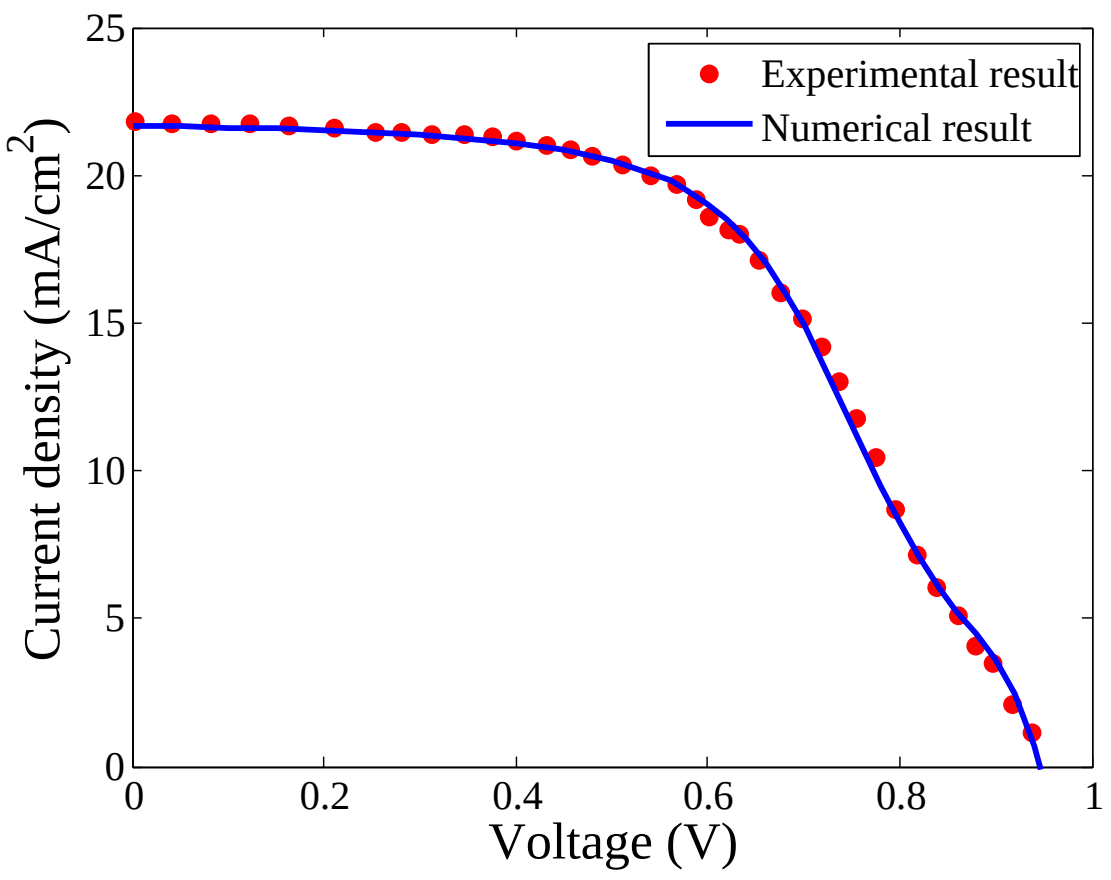

FIG. 2. Comparisons between experimental and numerical results. The experimental result is extracted from figure 4(b) in the reference 23.

## III. Simulation results and discussion

### A. The effect of light intensities

We first explore the effect of the different recombination channels on the performance of PSCs, such as *FF*, under various light intensities. The interfacial trap density for the top and bottom cases is $N_t = 6 \times 10^{15} cm^{-3}$, and the corresponding bulk trap density is $6.36 \times 10^{13} cm^{-3}$. As shown in Fig. 3(a), due to the thinner interfacial recombination region, the *FF* (above 80%) of PSCS with trap-assisted recombination at the top interface is larger than that (75%) of PSCs with trap-assisted recombination in bulk. Interestingly, the light intensity dependence of *FF* of PSCs with trap-assisted recombination at the bottom interface is anomalous when the light intensity is around 0.1 Sun. To reveal the cause of the anomalous *FF*, the *J-V* characteristics of PSCs at different recombination position with light intensity of 0.1 Sun are produced for comparison as shown in Fig. 3(b). Compared with the PSCs with the bulk recombination and the top interfacial recombination, an anomalous *J-V* curve of the PSCs with the bottom interfacial recombination, where the gradient of the *J-V* curve is changed obviously at the kink point when the applied voltage is larger than 0.7 V, was observed. As shown in Fig. 3(a), it is noted that there is a minimum of *FF* of PSCs in the case of bottom interfacial recombination when the light intensity is 0.1 Sun. To investigate the anomalous phenomenon, Figure 3(c) shows the *J-V* characteristics of PSCs with light intensity of 0.01, 0.1 and 1 Sun for the case of bottom interfacial recombination. The applied voltages corresponding to the occurring maximum bending points of *J-V* curves are equal (0.83 V) when the light intensity is above 0.01 Sun. For light intensity is lower than 0.01 Sun, $V_{OC}$ is lower than 0.8 V and the *FF* does not decrease due to no bending of *J-V* curves. For light intensity is 0.1 Sun, the *FF* decreases to the minimum because it is just near the maximum power point (MPP) of the PSCs.

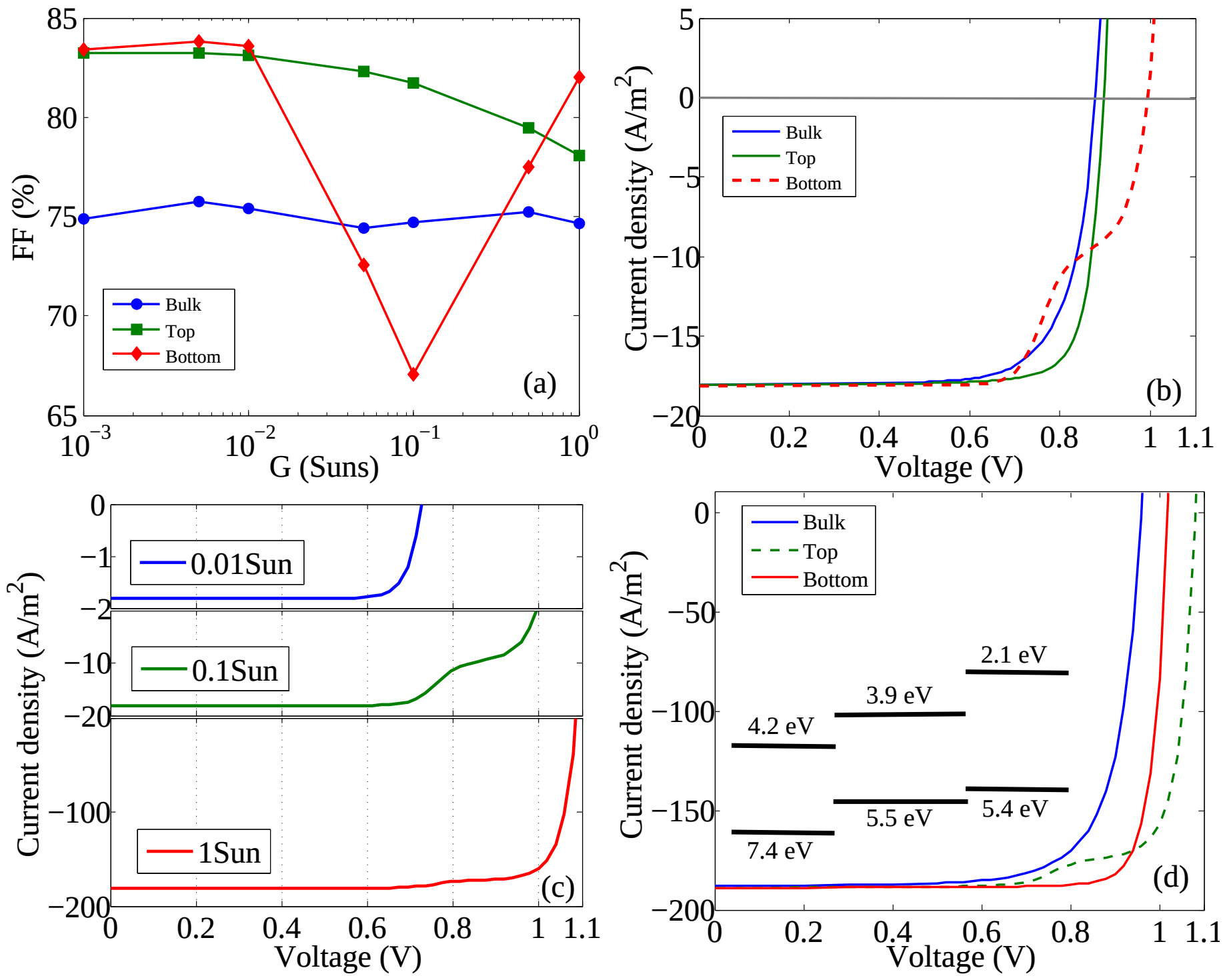

FIG. 3. (a) Light intensity dependence of *FF* for proposed PSCs with interfacial trap density $6 \times 10^{15} cm^{-3}$, and the corresponding bulk trap density $6.36 \times 10^{13} cm^{-3}$. (b) The *J-V* characteristics of PSCs with different trap-assisted recombination channel for light intensity of 0.1 Sun. (c) The *J-V* characteristics of PSCs with light intensity of 0.01, 0.1 and 1 Sun for the case of bottom interfacial recombination. (d) The *J-V* characteristics of PSCs with exchanged injection barrier (inset) for light intensity of 1 Sun.

However, it is worth our consideration, with equal trap density at top interface, no similar anomalous *J-V* curves are observed. As shown in Fig. 1(b), the injection barrier of electron at the ETL/perovskite interface (0.1 eV) and that of hole at perovskite/HTL interface (0.3 eV) is unbalance. The injection barrier of hole at the bottom is larger. If we exchange the injection barrier of each other, conversely, the *J-V* curve bending will exist in the case of top interfacial recombination and disappear in the case of bottom interfacial recombination as shown in Fig. 3(d). The anomalous *J-V* characteristics of PSCs originate from a combination of large enough injection barrier and interfacial recombination at the same interface. The underlying mechanism will be described in the following sections.

**B. The effect of trap densities**

The trap-assisted recombination is the dominant recombination mechanism during device operation[27]. Moreover, the impact of recombination at different positions on the device performance is also various, particularly with higher trap density. $V_{OC}$ decreases with increasing trap density due to increasing trap-assisted recombination at different recombination location. Figure 4(a) shows the *FF* of PSCs with trap density from $1 \times 10^{14}$ to $3 \times 10^{16} cm^{-3}$. *FF* is sensitive to the recombination location and strength in the device. The *FF* of PSCs with bulk recombination is obviously lower than that with top interfacial recombination when $N_t$ is above $6 \times 10^{14} cm^{-3}$. So, the quality of perovskite film should be enhanced together with passivation of traps at top interface for achieving higher performance of PSCs. The *FF* of PSCs with bottom interfacial recombination drops rapidly when $N_t$ is above $6 \times 10^{14} cm^{-3}$. To analyze the impact of bottom interfacial recombination and find out the physical origin, *J-V* characteristics of PSCs for bottom interfacial recombination with trap density from $1 \times 10^{14}$ to $3 \times 10^{16} cm^{-3}$ are illustrated in Fig. 4(b). The *J-V* curve begins to bend upwards when $N_t$ is above $4 \times 10^{15} cm^{-3}$.

And the bigger the trap density, the stronger the *J-V* curves bend. As can be seen from the figure, the voltage corresponding to the maximum degree of bending is the same, i.e., 0.83 V.

A voltage applied at the electrodes of the PSCs establishes an electric field within the device that forces the carriers to move with an average drift velocity. Reduction of the electric field intensity results in slower extraction of charge carriers in the bottom interfacial recombination region, leading to higher carrier recombination. As shown in Fig. 4(c), with the trap density $N_t$ $3 \times 10^{16} cm^{-3}$, potential profile at applied voltage (0 V) for short circuit is linear dependence on position in perovskite. The current density is max at short circuit due to large electric field. When the applied voltage is 0.83 V where the maximum bending of *J-V* curve exists, the potential is constant in the bottom interfacial region. So a zero electric field is generated there, resulting in a deep drop of current density due to increasing the interfacial recombination. The bottom interfacial recombination rate profiles of PSCs with various trap densities are illustrated in Fig. S2 in the supplementary material. The higher trap densities result in the more recombination at the interfacial recombination region. At open circuit, the potential is almost constant in the bulk perovskite, except in the bottom interfacial region, resulting in net zero current density because of small electric field. For clearly illustrating the anomalous *J-V* characteristics occurring in specific PSCs with stronger bottom trap-assisted recombination with the trap density $N_t$ $3 \times 10^{16} cm^{-3}$, Fig. 4(d) shows the electric field trends in bulk region, as well as around top and bottom interfaces versus applied voltage. In the cases of bulk and top interfacial recombination, the zero electric field exists at 1.02 V which is near to open circuit voltage (1.05 V), so no obviously anomalous *J-V* bending exists. In the case of bottom interfacial recombination which lies from 515 to 520 nm in our model, the zero electric field exists at 0.83 V. The zero electric field profile at the bottom interface shows in the inset of Fig. 4(d).

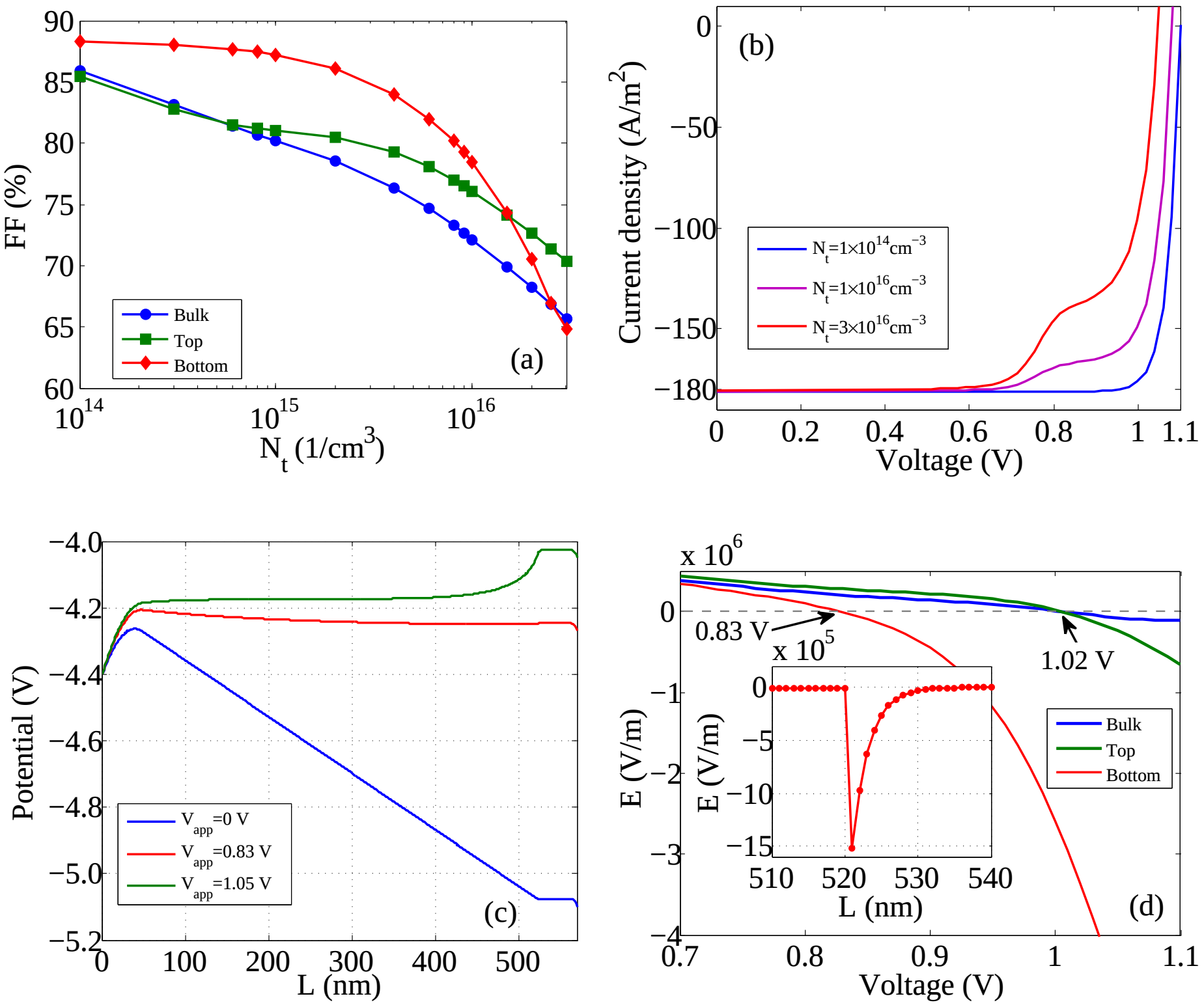

FIG. 4. (a) $FF$ of PSCs with trap density from$1\times10^{14}$ to $3\times10^{16}cm^{-3}$ at different recombination location. (b) J-V characteristics of PSCs for bottom interfacial recombination with various trap density. (c) Potential profiles at specific applied voltage, 0 V for short circuit, 0.83 V for 'max bending' of *J-V* curves and 1.05 V for open circuit with the trap density $N_t$ $3\times10^{16}cm^{-3}$. (d) Electric field in bulk, top and bottom interfacial region versus applied voltage with the trap density $N_t$ $3\times10^{16}cm^{-3}$, the inset graph is electric field profile for bottom interfacial region at 0.83 V.

## C. The impacts of doping concentration and injection barrier

The doping of charge carrier transport layer could be an effective method to increase the *FF* of PSCs[28] which benefits the faster charge extraction and thus resulting in lower recombination. The electric field in the charge carrier transport layer increases with the corresponding increasing the doping concentration. In other words, as shown in Fig. 5(b), the electrical potential at the perovskite/HTL interface decreases when increasing the doping concentration in HTL. The black dotted line in this figure represents the perovskite/HTL interface. Just for this reason, the applied voltage corresponding to the maximum degree of bending increases with increasing the doping concentration in HTL, as shown in Fig. 5(a). Besides, the current density is also sensitive to the trap density at the bottom interface region. For the higher doping concentration ($N_A^- = 1 \times 10^{18} cm^{-3}$), the *J-V* curve can also be distorted even if with the lower trap density ($N_t = 5 \times 10^{14} cm^{-3}$). As shown in Fig. 5(c), the distortion of *J-V* curve is more serious for those with higher doping concentration in HTL but lower trap density at bottom interface. *FF* of PSCs could be improved via appropriately increasing the doping concentration of charge carrier transport layer. It should be noted that, however, the *J-V* curve could be distorted with relatively lower trap density for higher doping concentration. There is a trade-off between doping concentration of charge carrier transport layer and trap density at the interface. Therefore, we should pay more attention to the passivation of the interface for PSCs with higher doping concentration in charge carrier transport layer to enhance the electric performance of PSCs.

Until now, we know that the degree of *J-V* curve bending, which impacts on the *FF* of PSCs, depends on different recombination channels, trap density at the interface and doping concentration in charge carrier transport layer. By investigating the profiles of charge carriers in the perovskite active layer of PSCs, we find that the injection barrier of the interface is the most fundamental reason resulting in the above anomalous *J-V* characteristics. Figure 5(d) shows the charge carrier profiles with 1 V applied voltage in the vicinity of interface for clarity. $\Delta E_C$ is the electron injection barrier at the ETL/perovskite interface, and $\Delta E_V$ is the hole injection barrier at the perovskite/HTL interface. For the case of lower carrier injection barrier, such as $\Delta E_C = \Delta E_V = 0.1 eV$, there is no accumulation of both electrons and holes at respective interface with 1 V applied voltage. For the case of higher injection barrier, such as $\Delta E_C = \Delta E_V = 0.3 eV$ here, there is more electrons (holes) accumulation at the perovskite/HTL (ETL/perovskite) interface because it is difficult for holes (electrons) to inject into the active layer from external circuit. For a particular injection barrier, the greater the applied voltage, the larger the carriers accumulate. As a result, the more carriers' accumulation, the more recombination between carriers and traps at the interfaces occurs. And it results in the distortion in *J-V* curve. The sufficient electron/hole injection barrier and interfacial recombination could give rise to the local kink of *J-V* curve. Figure S3 in the supplementary material shows the effect of the hole injection barrier between the perovskite layer and HTL on the applied voltage forming the zero electric field at the interface. As shown in Fig. S3(a), the distortion of *J-V* curve is more serious for those with lower injection barrier but equal trap density at bottom interface ($3 \times 10^{16} cm^{-3}$). It results from the more recombination between the accumulated electrons and traps at the interface, due to the greater applied voltage point which forms the zero electric field at the interface when the lower injection barrier, shown in Fig. S3(b).

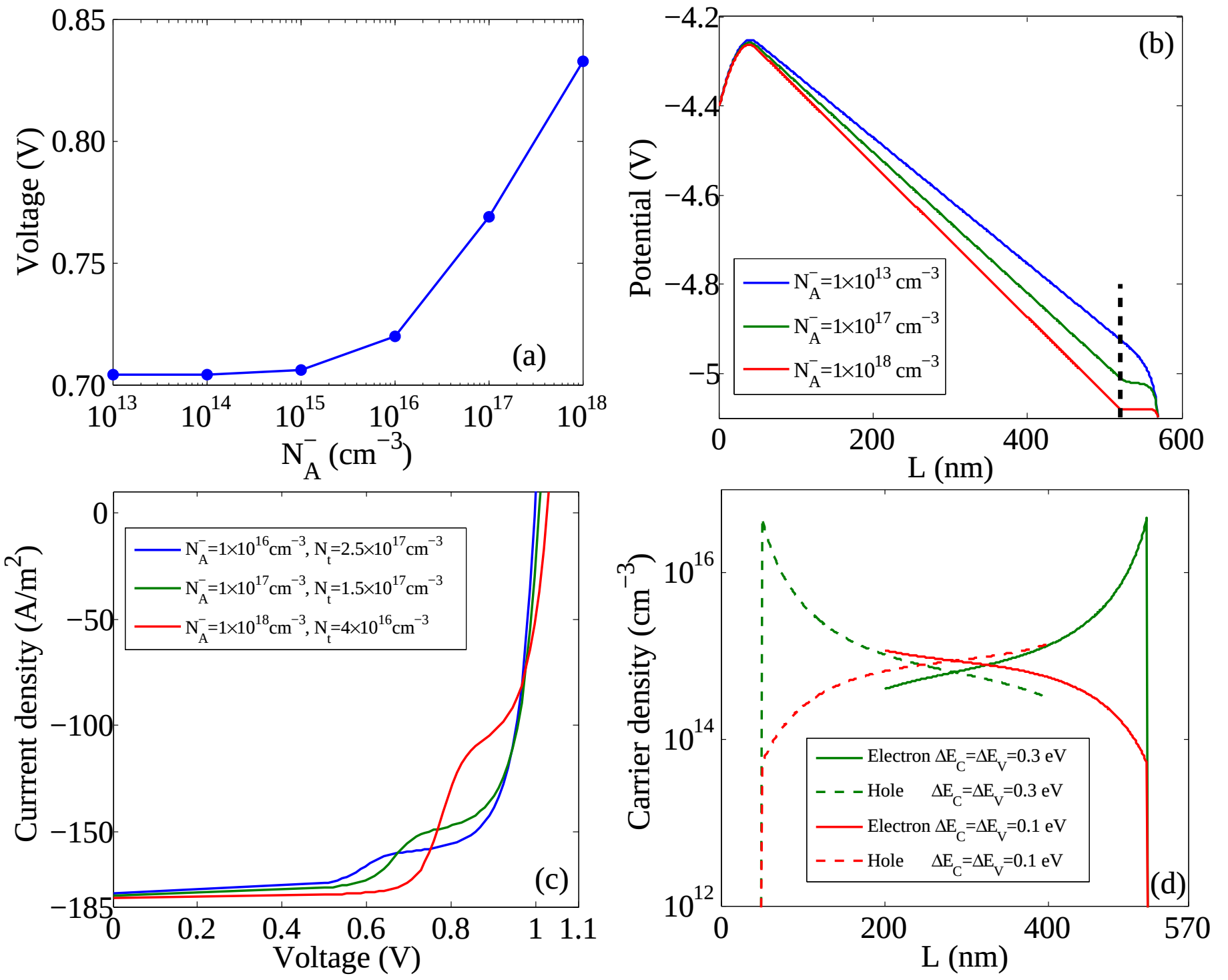

FIG. 5. (a) The voltage corresponding to zero electric field dependence of doping concentration in HTL. (b) Potential profiles at various doping concentration in HTL. (c) *J-V* characteristics of PSCs for bottom interfacial recombination with various trap density and doping concentration in HTL. (d) Charge carrier accumulation at respective interface with 1 V applied voltage, electrons accumulate at perovskite/HTL interface and holes accumulate at ETL/perovskite interface. The charge carrier profiles are only shown at the accumulation region for clarity.

## IV. Conclusions

We investigated the effect of trap-assisted interfacial recombination on the performance of PSCs based on the non-linear Poisson and drift-diffusion equations for electrons and holes throughout the device in one dimension. An anomalous *J-V* characteristic is observed in the study, and its origin is demonstrated to be the combination of large enough injection barrier and interfacial recombination. The injection barrier leads to the charge carrier accumulation in the interfacial region, which increases the trap-assistant recombination, resulting in a deep drop of current density at the specific voltage like 0.83V. It provides a basic routine to optimize the efficiency of PSCs in combination with energy band structure, doping concentration of the charge carrier transport layer and interface quality.

## Supplementary material

See supplementary material for the details of the computational method and additional simulation results.

## Acknowledgments

The project was supported by Natural Science Foundation of Jiangsu Province (Grant No. BK20140448 and BK20151284), Natural Science Foundation of the Jiangsu Higher Education Institutions of China (Grant No. 14KJD480001), National Natural Science Foundation of China (No. 61774069), the Innovation Foundation of HHIT (Grant No.Z2014018) and "Overseas training program for outstanding young teachers of universities in Jiangsu Province (2016)".

# Supplementary Material

A. Computational method used in solving the non-linear Poisson and drift-diffusion equations

In the model, we used the following equations:

$$\frac{\partial^2 V}{\partial x^2} = -\frac{q}{\varepsilon}(p - n - N_A^- + N_D^+) \tag{S1}$$

$$\frac{\partial n}{\partial t} = \frac{\partial}{\partial x}\left(\mu_n n E + D_n \frac{\partial n}{\partial x}\right) + G_n - R_n \tag{S2}$$

$$\frac{\partial p}{\partial t} = -\frac{\partial}{\partial x}\left(\mu_p p E - D_p \frac{\partial p}{\partial x}\right) + G_p - R_p \tag{S3}$$

where the physical explanations of all parameters are described in the manuscript. To solve above equations, the Scharfetter-Gummel scheme in the spatial domain and the semi-implicit strategy in the temporal domain are used[1–3]. The one-dimensional discretized forms of Eqs. (S1) - (S3) are respectively given by:

$$\frac{1}{\Delta x^2}\varepsilon_{i+\frac{1}{2}} V_{i+1}^{m+1} - \left[\frac{2}{\Delta x^2}\left(\varepsilon_{i+\frac{1}{2}} + \varepsilon_{i-\frac{1}{2}}\right) + \frac{n_i^m + p_i^m}{V_t}\right] V_i^{m+1} + \frac{1}{\Delta x^2}\varepsilon_{i-\frac{1}{2}} V_{i-1}^{m+1} = -q\left(p_i^m - n_i^m - {N_A}_i^m + {N_D}_i^m\right) - \frac{n_i^m + p_i^m}{V_t} V_i^m \tag{S4}$$

$$\left[\frac{1}{\Delta t} + D_{n,i+\frac{1}{2}} B\left(\frac{V_i^{m+1} - V_{i+1}^{m+1}}{V_t}\right) + D_{n,i-\frac{1}{2}} B\left(\frac{V_i^{m+1} - V_{i-1}^{m+1}}{V_t}\right)\right] n_i^{m+1} - D_{n,i+\frac{1}{2}} B\left(\frac{V_{i+1}^{m+1} - V_i^{m+1}}{V_t}\right) n_{i+1}^{m+1} - D_{n,i-\frac{1}{2}} B\left(\frac{V_{i-1}^{m+1} - V_i^{m+1}}{V_t}\right) n_{i-1}^{m+1} = \frac{1}{\Delta t} n_i^m + G_n^m - R_n(n_i^m, p_i^m) \tag{S5}$$

$$\left[\frac{1}{\Delta t} + D_{p,i+\frac{1}{2}} B\left(-\frac{V_i^{m+1} - V_{i+1}^{m+1}}{V_t}\right) + D_{p,i-\frac{1}{2}} B\left(-\frac{V_i^{m+1} - V_{i-1}^{m+1}}{V_t}\right)\right] p_i^{m+1} - D_{p,i+\frac{1}{2}} B\left(-\frac{V_{i+1}^{m+1} - V_i^{m+1}}{V_t}\right) p_{i+1}^{m+1} - D_{p,i-\frac{1}{2}} B\left(-\frac{V_{i-1}^{m+1} - V_i^{m+1}}{V_t}\right) p_{i-1}^{m+1} = \frac{1}{\Delta t} p_i^m + G_n^m - R_n(n_i^m, p_i^m) \tag{S6}$$

where the superscript $m$ and subscript $i$ indicate the discretized temporal and spatial steps, respectively. $V_t = \frac{k_B T}{q}$ is the thermal voltage and $B(x) = \frac{x}{e^x - 1}$ is the Bernauli's function.

The grid spacing is 1 nm. The convergence condition for the steady solution calculation is given by $\left|\frac{J^{m+1} - J^m}{J^m}\right| < 10^{-10}$.

B. Charge carrier generation profile throughout the device

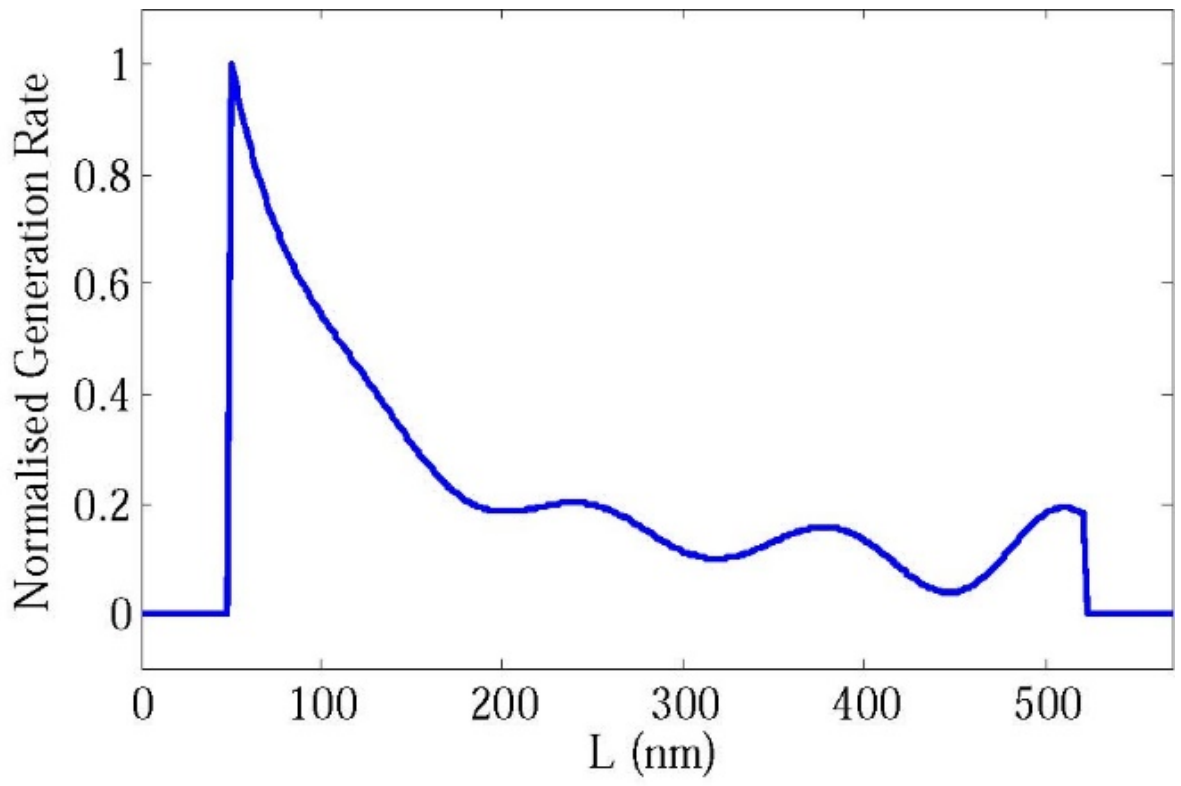

FIG. S1. Normalized generation rate of charge carriers throughout the device.

C. Additional simulation results

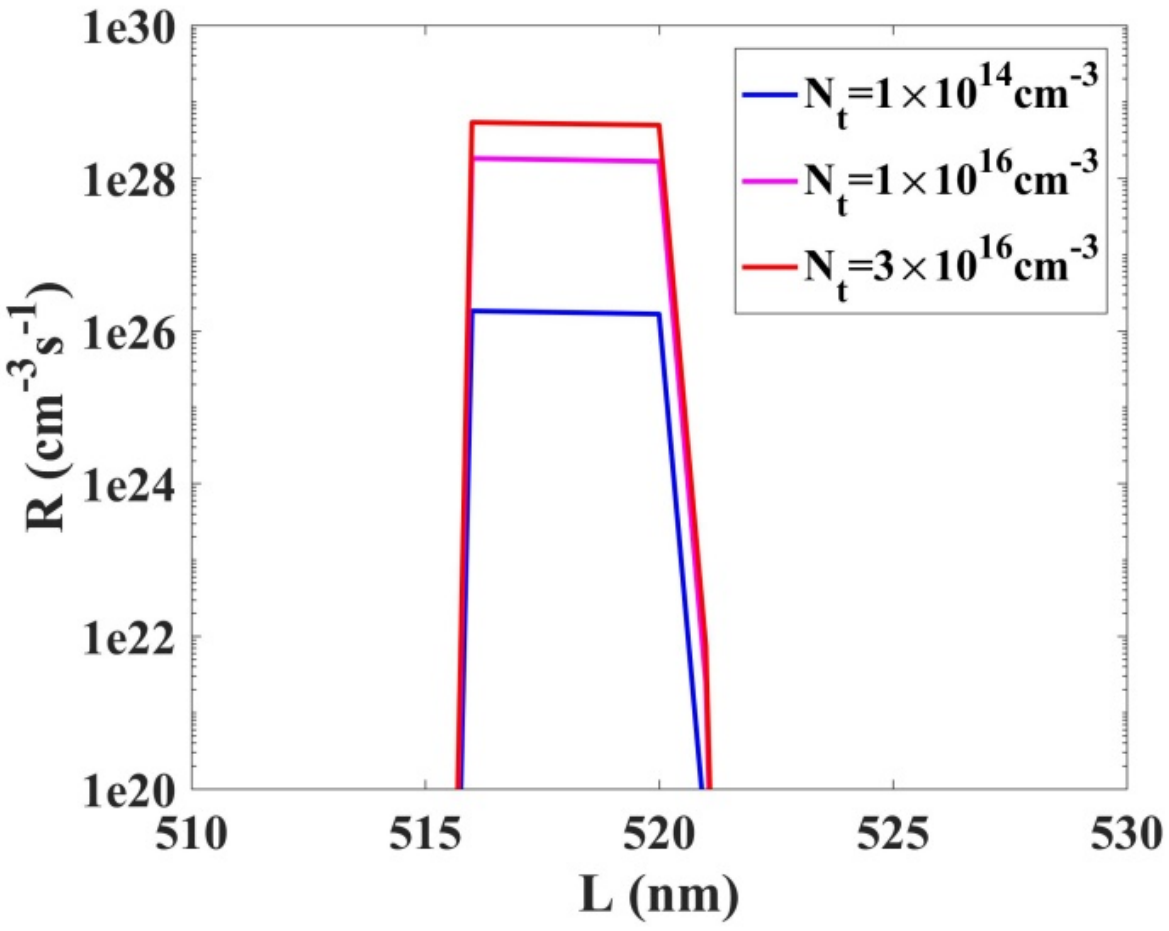

FIG. S2. The interfacial recombination rate profiles of PSCs with trap density from$1 \times 10^{14}$ to $3 \times 10^{16} cm^{-3}$ in the bottom interfacial recombination region.

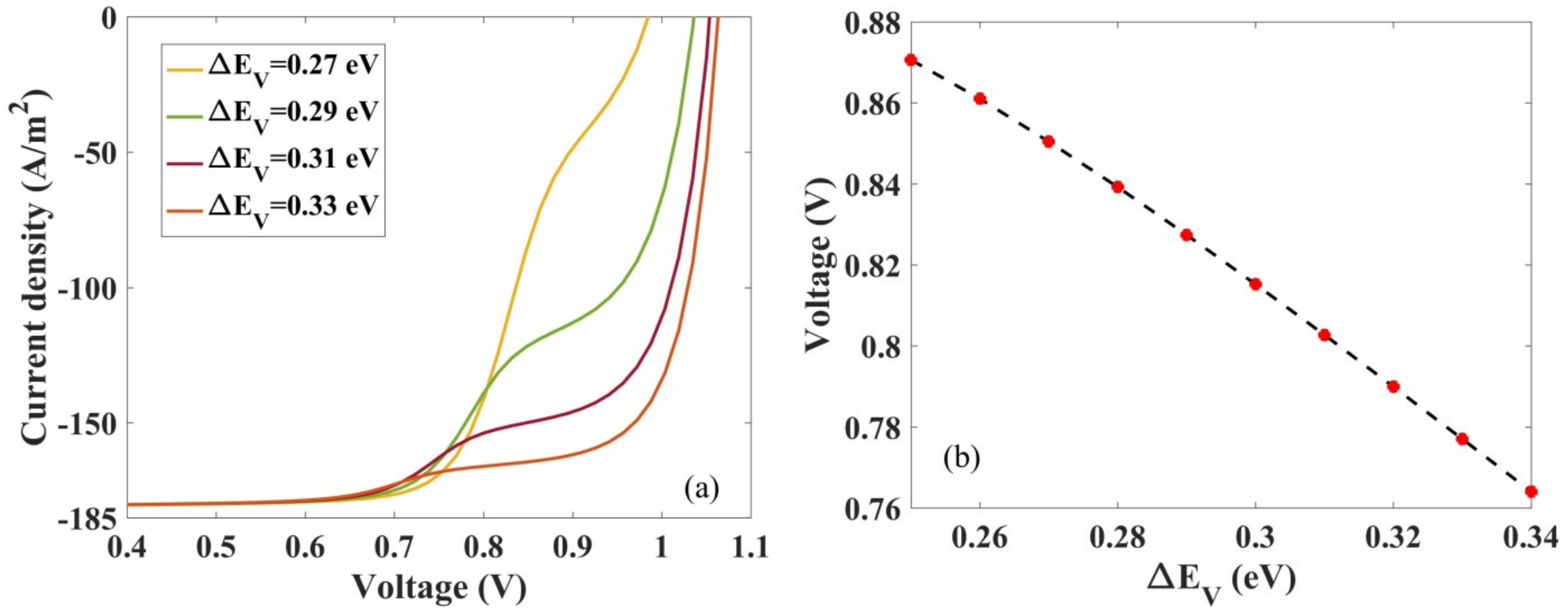

FIG. S3. The effect of the hole injection barrier between the perovskite layer and HTL on the applied voltage forming the zero electric field at the interface. (a) *J-V* characteristics of PSCs with various hole injection barrier. The trap density at the bottom interface is $3 \times 10^{16} cm^{-3}$. (b)

The applied voltage corresponding to the zero electric field dependence of the hole injection barrier.